\documentstyle[12pt, epsfig] {article}
\begin {document}
\title
{REALISTIC EXPERIMENTS FOR MEASURING THE WAVE FUNCTION OF A SINGLE PARTICLE}
\author {S. Nussinov \\
School of Physics and Astronomy\\
Raymond and Beverly Sackler Faculty of Exact Sciences\\
Tel Aviv University,  Tel Aviv 69978 Israel\\
and\\
University of South Carolina\\
Columbia, South Carolina 29208
USA}
\maketitle
\begin {abstract}
\par
We suggest scattering experiments which implement the concept of
\newline
``protective measurements''  allowing the measurement of the complete wave
function even when only one quantum system (rather than an ensemble) is
available.  Such scattering experiments require massive, slow, projectiles
with kinetic energies lower than the first excitation of the system in
question.  The results of such experiments can have a (probabilistic)
distribution (as is the case when the Born approximation for the scattering
is valid) or be deterministic (in a semi-classical limit).
\end {abstract}
\newpage

{\bf I.     Introduction.}

\par
Since the very early days of quantum mechanics, it is generally
believed$^{1}$ that the wave function of a single quantum system, say, a
hydrogen atom in its ground state, is not a directly measurable quantity.
Rather $\mid \psi (x) \mid^{2}$ dV is simply the probability that the
electron will be found within a volume element dV around the point $x$.
Since localization of the electron does destroy the wave function, we need
an ensemble of many identical quantum systems - say many hydrogen atoms in
the ground state - to measure the wave function.  The latter is therefore
believed to be only a mathematical symbol devoid of direct physical
reality$^{2}$ at the one-system (atom) level.
Recently,$^{3}$ it has been pointed out, however, that the standard
framework of quantum mechanics does allow measuring the wave function of a
single particle in a bound, stationary, state.  The ``protective''
measurements required for this purpose involve weak, adiabatic interactions
of the particle with the measuring device. 
This suggestions was discussed quite extensively in the literature. 
Refs. 4-6 contain some critical comments which were answered in
Ref.7. Refs. 8-10 address peculiar features of the proposed novel
measurements. Recent further elaboration of the basic idea in the
context of the two-state vector formalism$^{11}$ were presented in Refs.12,13.

 In the protective measurements the
``pointer'' indicates not an eigenvalue, $a$, of the measured operator $A$,
as is in the case in the standard `impulsive' measurements, but rather the
expectation value:
\begin {equation}
\bar{A} = \langle \psi_{0} \mid A \mid \psi_{0}\rangle 
\end {equation}
of $A$ in the state $|\Psi_0 \rangle$ of interest.
\par In the  case of impulsive measurements the initial wave function
$\mid \psi_{0}\rangle $ is reduced (``collapsed'') into one of the eigenstates $
\mid a \rangle $ of the operator $A$, specifically to the state whose eigenvalue
 is indicated by the pointer. The probability of a specific outcome, $a$,
is:
\begin {equation}
P(a) =\mid\langle a\mid \psi_{0}\rangle \mid^{2}
\end {equation}
Thus if $A$ is the position operator $X$ the probability of finding
particle at $X = x$ is
\begin {equation}
\mid \langle  x \mid \psi_{0} \rangle  \mid^{2} \equiv \mid \psi_{0} (x) \mid^{2}
\end {equation}
with $ \psi_{0} (x) $ the wave function in the usual coordinate basis.
\par The recently suggested protective measurements are radically different:
\newline (a) First, after the protective measurement of the operator
$A$ is completed, the particle stays in the same initial state
$\psi_{0}$ with a probability that can be made arbitrarily close to 1.
Specifically, the excitation to other states, assuming these are at a
minimal (energy) distance $\Delta E$ away, can be made exponentially
small$^{14}$
\begin {equation}
P_{exc} \approx e^{-\Delta E \cdot T}
\end {equation}
with $T$ the duration of the adiabatically switched interaction.
\newline
(b) Second, the result of the measurement, i. e. the pointer position is
deterministic (rather than probabilistic) and equals the average of $A$ in
the state  $\psi_{0}$.  The accuracy of this last statement depends on the
validity of first order perturbation in the interaction $V_{I}$ coupling
the particle and measuring device, i.e. on the assumption that during the
measurement process the wave function is only weakly distorted.
\par Using feature (a) we can repeat the protective measurement for N
different operators $A_{i}$.  Knowledge of $\bar{A}_{i} = \langle  \psi_{0} \mid
A_{i} \mid \psi_{0} \rangle $ allows inferring the initial wave function with
precision $O (1 / N)$.  We also need $N$ impulsive, position, measurements
in order to sample the probability $ P(x) = \psi_{0}^{2} (x)$ at $N$ points
$x_{i}$, $i = 1, ..., N$ and get $\psi_{0}^{2} (x)$ to a requisite $O (1 /
N)$ precision.  There is, however, one crucial difference: Our original
wave function is not destroyed by the repeated protective measurements so
that {\it one} bound electron is all that is needed.
\par
The purpose of the present paper is two-fold.  First, we would like to
point out (almost) ``real experiments'' in which we can, by repeated
scattering on the same quantum system, say hydrogen in the ground state, map
out the complete wave function, without ever ionizing or exciting the
system.  This is facilitated simply by using massive slow projectiles such
that on the one hand their momentum is large enough to allow probing the
atomic wave-function of interest and on the other their kinetic energy is
smaller than the energy gap $\Delta E$ between the ground state and the
first excited state.  Our second result is that the two aspects (a) and (b)
of the recently suggested protective measurements-namely the
non-destruction of the wave function and the deterministic rather than
probabilistic result of the measurements - are really independent.  Thus we
could have scattering experiments involving heavy projectiles which can
never destroy the state of interest, i. e. which are completely elastic,
where the elastic scattering is probabilistic.  This in particular is the case
when the Born approximation is applicable.  The result of individual
scattering could also be deterministic - e.g., when an ``eikonal
approximation'' involving the exchange of many quanta is applicable.
\par The plan of the paper is the following.  In Section II we briefly
review the protective measurements of Aharonov, Anandan, and Vaidman (Ref.
3).  Section III recalls the general connection between measurements and
scattering experiments.  It also explains why generically such experiments
are of the ``impulsive'' type and tend to break the bound state.  Section
IV contains our main result namely the description of scattering
experiments involving heavy projectiles so that they do not excite the
target yet map out the wave function.  For concreteness we assume that the
scattering can be described in Born approximation.  Section V extends the
discussion to a case where many quanta are exchanged in typical scattering.
The resultant deflection is then (for a given initial impact parameter
$b$) practically deterministic.  This corresponds more closely to the
originally suggested scheme of Aharonov and Vaidman and Aharonov, Anandan
and Vaidman.$^3$
\par
Finally, in the summary in Section VI, we speculate on {\it real}
experiments and discuss various putative heavy projectiles, which can
implement these experiments.


{\bf II.     The General von Neumann Formalism and the Novel Protective
Measurements.}

\par
Let us first briefly review the basic von Neumann formalism$^{15}$ underlying
all quantum measurements, impulsive and protective, alike.  The system of a
particle of mass $m$, coordinate $\vec {r}$, momentum $\vec{p}$ and the
measurement apparatus (``pointer'') with coordinate $Q$, momentum $P$ and
mass $M \gg m$ are initially non-interacting.  The Hamiltonian then
separates into a sum of two parts pertaining to the particle and pointer
respectively:
\begin {equation}
H_{0} = [ \frac {\vec{p}^{2}} {2m} + V_{B} (\vec {r}) ] + [ \frac
{\vec{P}^{2}} {2M} + V (\vec{Q})]
\end {equation}
with $V_{B} (\vec{r})$, $V(\vec{Q})$ binding potentials.  The corresponding initial
state factorizes into two parts.
The particle is assumed to be initially in a state $ \mid \psi_{0} \rangle  $
which is a stationary bound state, say the ground state, in the potential
$V_{B} (\vec{r})$.  The pointer is assumed to be initially in some wave
packet state, say a Gaussian: $\Phi_{0}(Q)= exp[-\frac{Q^{2}}{2(\Delta
Q)^{2}} ]$ centered around the origin in coordinate space [or alternatively
as the Fourier transformed Gaussian $\tilde{\Phi}_0 (P) = exp [- \frac {P^{2}}
{2(\Delta P)^{2}}]$ in momentum space, with $\Delta P\Delta Q = \hbar/2.$].
The measurement process of an operator $A(\vec{r}, \vec{p})$ consists of
turning on, an interaction of the form:
\begin {equation}
H_{I}(t) = g(t) A(\vec{r}, \vec{p}) \cdot P.
\end {equation}
between the system and pointer.  The time profile of switching the
interaction on and off is normalized by
\begin {equation}
\int_{-\infty}^{\infty} g(t) dt = 1.
\end {equation}
In the impulsive measurements $g(t)$ differs from zero only over a short
time interval $\delta$.  During this time
$ H_{I}\approx \delta^{-1} $
is very large and we can neglect $H_{0}$ in evaluating the evolution of the
system
\begin {equation}
U(t, t+\delta) = exp [ - i \int_{t}^{t+\delta} g(t) A(\vec{r}, \vec{p}) \cdot P].
\end {equation}
The initial state of the particle is decomposed in the basis of $A$
eigenstates: \newline $\mid\psi_{0}\rangle  = \sum \mid a \rangle \langle  a\mid \psi_{0} \rangle$.
Upon applying $U(t+\delta,t)$ each component $\mid a\rangle$ acquires the
appropriate phase:
\begin {equation}
\begin{array} {c}
U(t+\delta,t) \{\mid \psi_{0} \rangle \times \mid \Phi_{0} \rangle\} = \sum_{a} e^{iaP}
\mid \Phi_{0} \rangle \mid a \rangle \langle  a \mid \psi_{0} \rangle
\\
= \sum_{a} \mid a\rangle \mid \Phi (a) \rangle \langle  a \mid \psi_{0} \rangle
\end {array}
\end {equation}
where $\mid\Psi (a) \rangle$ is the pointer wave function shifted via the
$e^{iaP}$ translation operator from the initial position around $Q = 0$, to
a similar Gaussian packet centered around $Q = a$ (so that the pointer
indicates the specific eigenvalue $a$ measured).
\par
Let us next consider the protective measurements suggest by AV and
AAV.$^{3}$  In these measurements $g(t)$ is turned on for a long time $T$.
Eq. (7) implies then a very weak $H_{I}$: $g(t) \approx \frac {1}{T} =
\epsilon$.  If the turning on and off of the interaction is sufficiently
smooth then we can use
\par (i) the adiabatic approximation for treating the effect of $H_{I}$.
The weakness of $H_{I}$ allows us, furthermore, to do this \par (ii) in a
first order perturbation for $H_{I}$.  \par Thus (i) implies that the
original state $\psi_{0}$ gets continuously deformed to $\psi_{0} (t)$ with
$\psi_{0}(t)$ being the instantaneous eigenfunction of the time-dependent
Hamiltonian $H(t) = H_{0} = H_{0} + H_{I}(t)$ with eigenvalue $E_{0}(t)$:
\begin {equation}
H(t) \mid \psi_{0}(t) \rangle = E_{0}(t)\mid\psi_{0}(t) \rangle
\end {equation}
and we have:
\begin {equation}
\psi_{0}(t) \rightarrow_{t\rightarrow \pm \infty} \psi_{0};~~~~
E_{0}(t) \rightarrow_{t\rightarrow \pm \infty} E_{0}.
\end {equation}
The probability of leaving the system in some excited final state is
exponentially small:$^{14}$
\begin {equation}
P_{exc} \approx e^{-\Delta E\cdot T}.
\end {equation}
\par The second feature implies (ii) that the energy shift $\delta E_{0}
(t) = E_{0}(t) - E_{0}$ can be evaluated by first order perturbation in the
weak interaction $H_{I}$:
\begin {equation}
\delta E_{0}(t) = \langle  \psi_{0} \mid H_{I}(t) \mid \psi_{0} \rangle = g(t) P \langle 
\psi_{0} \mid A \mid \psi_{0} \rangle \equiv g(t) P \bar{A}
\end {equation}
The corresponding time evolution of the pointer will be given by
\begin {equation}
e^{-i\int \delta E_{0}(t)dt} \mid Q_{0} \rangle = e ^{-i \int dtg(t)P \bar{A}} \mid
Q_{0} \rangle = \mid Q_{0} + \bar{A} \rangle
\end {equation}
i. e. the Gaussian wave packet of the pointer will shift by
\newline
$\bar{A} \equiv \langle  \psi_{0} \mid A \mid \psi_{0} \rangle$.  Such a shift can be
discerned if it is larger than the original width (in configuration space)
of the $Q$ wave packet, i. e. if
\begin {equation}
\bar{A} \geq (\Delta Q).
\end {equation}
Using Eq. (12), the fact that the transition frequencies to all the higher
levels exceed $\Delta E$, and completeness:
\begin {equation}
\sum_{n} \mid \langle  \psi_{n} \mid A \mid\psi_{0} \rangle \mid^{2} = \langle  \psi_{0} \mid
A^{2} \mid \psi_{0} \rangle
\end {equation}
it can be shown that the sum of all excitation probabilities to any bound (or unbound) excited state satisfies
\begin {equation}
\sum_{n \neq 0} P (\psi_{0} \rightarrow \psi_{n}) < C \cdot
e^{-\Delta ET}.
\end {equation}
Thus such a measurement indeed ``protects'' the initial state and maintains
it for subsequent experiments of the same protective nature, where other
operators
\newline
$A_{i}$, $ i = 1, ..., N$ are measured.
\par In particular, AV and AAV$^{3}$ choose the operators $A_{i}$ to be
$P(\delta v_{i})$, the projections onto many small space regions around
points $\vec{r}_{i}$.  Since the ground state wave function is real and
nodeless, measuring
$\langle  P\delta v_{i} \rangle = \int _{\delta v_{i}} \psi_{0}^{2} (\vec{r})$
fixes $\psi_{0} (\vec{r}_{i})$.  Letting $N \rightarrow  \infty$
recuperates the complete wave function.
\newline

{\bf III.     Measurement as Scattering Experiments.}

The formal von Neumann pointer-system interaction is similar to usual
scattering.  Let us scatter form the electron, bound to infinitely heavy
``proton'' $P$, at the origin, a projectile particle, $X$.  The latter
is initially
moving with velocity $V$ along the $z$ axis, say.  The scattered projectile
particle plays here the role of the pointer.  The interaction shifts its
momentum by $\Delta \vec{P}$ and deflects it by an angle
$ \Delta \theta \approx \mid \Delta \vec{P} \mid /P$.
This deflection is finally translated, in the subsequent free evolution of
the scattered wave packet, into a lateral displacement on a far-away
screen.
\par
 A large class of $\gamma-ray$, neutron or electron scattering,
measure the
``Form Factor'' of quantum systems.  The differential scattering cross
section of electrons, say, in the Born approximation, is proportional to $
\mid F \mid ^{2}$ with  $F$ the ``Form Factor'':
\begin {equation}
F(\Delta \vec{P}) \approx \int d \vec{r}
e^{i \Delta \vec {P} \cdot \vec{r}} \rho (\vec {r})
\end {equation}
and $\rho$ the charge density.  If the charge density is contributed by one
electron in the $\psi_{0} (\vec{r})$ bound state, then $\rho(\vec{r})$ is
given by:
\begin {equation}
\rho(\vec {r}) = q \mid \psi_{0} (\vec{r}) \mid ^{2},
\end {equation}
Once $d \sigma / d \Omega$ is measured, we can infer $\mid F(\Delta
\vec{P}) \mid^{2}$.  For the case $\rho (\vec{r}) = \rho(-\vec{r})$, $F$ is
real and $F(\Delta \vec{P})$ can also be determined.  By (inverse) Fourier
transformation, we would then find $\psi_{0} (r).$  However, in general,
this measurement cannot be done by scattering on one single atom and an
ensemble of atoms is required.  This is related to the dual significance of
the form factor $F(\Delta \vec{P})$ in impulsive scattering; namely that
$\mid F (\Delta P)\mid^{2}$ is then also the probability of not destroying
the system in any, specific, scattering event.  Indeed, for collision time
short in comparison with atomic periods, the ``sudden'' approximation is
appropriate.  The transfer of momentum
$\Delta \vec{P} = \vec {P}_{f} - \vec{P}_{i}$
transforms the initial wave function via:
\begin {equation}
\psi_{0} \rightarrow \psi_0 '
= e^{i \Delta \vec{P} \cdot \vec{r}} \psi (\vec{r}).
\end {equation}
The probability of staying in the ground state is then the square of the
form factor:
\begin {equation}
P _{\psi_{0} \rightarrow \psi_0 '}
= \mid \langle  \psi_{0} \mid \psi_0 ' \rangle \mid ^{2}
= \mid \int d \vec{r}
 e^{i \Delta \vec{P} \cdot \vec {r}} \psi_{0} (r) \mid^{2}  = \mid F
(\Delta \vec{P}) \mid ^{2}
\end {equation}
The normalization condition $\int d \vec{r} \psi _{0}^{2}(r)$ = 1 implies
that $F(\Delta \vec{P} = 0) = 1$.  However,
\begin {equation}
\mid F [(\Delta \vec{P}) \neq 0] \mid <  1
\end {equation}
The probability that the initial atom stays
in the ground state after $N$ repeated scattering
with momentum transfers $\Delta \vec{P}_{1}$, $\Delta
\vec{P}_{2},...,\Delta \vec{P}_{i}..\Delta \vec{P}_{N}$, is therefore
exponentially small:
\begin {equation}
P_{\psi_{0}\rightarrow \psi_{0}} \mbox{(after N collisions)}
= \prod_{i=1}^{N}
 \mid F (\Delta \vec{P}_{i}) \mid^{2} \approx e^{-cN}.
\end {equation}
[To show this recall that the geometric mean of $N$ positive numbers is
smaller than the arithmetic mean.  Hence:
\begin{equation}
\prod _{i=1}^{N} F^{2} (\Delta \vec{P}_{i}) \leq [ \frac {1} {N}
\sum ^{N} F^{2} (\Delta \vec{P}_{i})]^{N}
\approx (\bar{F^{2}})^{N} \approx e^{-cN}
\end {equation}
with $\bar{F^{2}}$ an effective average of $F^{2}$ obtained by the
experimental sampling.]  Consequently, we need $\approx N$ atoms if we want
to measure the form factor at $N$ points.
\newline

{\bf IV.     ``Protective'' Scattering:  The Born Approximations}

\par
We have seen that typical elastic scattering experiments are similar to
impulsive von Neumann measurements.  We proceed next to discuss adiabatic
scattering corresponding to protective measurements.
\par If $\mid \Delta \vec{P} \mid \approx P sin \theta$ is small in comparison
with the initial momentum $\vec{P} = MV \hat{e}_{z}$ of the projectile,
then the deflection angle is small:
\begin {equation}
\theta = \mid \Delta \vec{P} \mid / P \approx
\frac {\mid \Delta \vec{P}\mid} {MV} \ll 1.
\end {equation}
The effective collision time, estimated via the classical traversal time,
is then:
\begin {equation}
\delta t _{col} \approx a_{0}/V
\end {equation}
with $a_{0}$ the effective size of the state $\psi_{0}$.  By the
uncertainty principle the average (rms) momentum of the electron satisfies
\begin {equation}
p \geq \hbar/a_{0}
\end {equation}
If $m$ is the electrons' mass, then:
\begin {equation}
v_{e} \equiv v \approx p/m \approx \frac {\hbar} {ma_{0}}
\end {equation}
\begin {equation}
\Delta E\approx\frac{mv^{2}}{2} =\frac{\hbar^{2}}{2ma_{0} ^{2}}
\end {equation}
\begin {equation}
t_{atom} \approx \frac {a_{0}}{v}
\approx \frac {ma_{0}^{2}}{\hbar}
\approx \frac{\hbar}{\Delta E}
\end {equation}
are the electron's ``velocity'', the energy difference to the first excited
state, and the atomic period respectively.  The sudden approximation used
above was predicated on the assumption that
\begin {equation}
\delta t_{col} \ll t_{atom}
\end {equation}
or, using Eqs. (30) and (26) on:
\begin {equation}
V \gg v \mbox{[sudden approximation]}.
\end {equation}
If, further, $M \geq m$, this implies a kinetic energy of the projectile,
$KE = \frac{1} {2} MV^{2}$, which considerable exceed $\Delta E \approx
\frac {1} {2} mv^{2}$ and the atom can be readily excited or ionized we are
however interested.
In the opposite, adiabatic case,
\begin {equation}
\delta t _{col} \gg t_{atom}
\end {equation}
or
\begin {equation}
v \gg V \mbox{[adiabatic approximation]}
\end {equation}
In order to be able to effectively probe $F(\Delta \vec{P})$ over the whole
relevant range of $\mid \Delta P \mid \approx \hbar/a \approx mv$, we need
that:
\begin {equation}
\mid \Delta \vec{P} \mid \approx MV \theta  > p \approx mv
\end {equation}
or
\begin {equation}
MV \gg mv
\end {equation}
\par
Clearly, Eq. (36) is consistent with Eq. (34), providing that the
projectile-electron mass ratio is large enough:
\begin {equation}
f \equiv M/m \gg 1.
\end {equation}
\par
Let us impose next a stronger version of the adiabatic condition namely $V
\leq v / \sqrt{f}$ so that
\begin {equation}
\frac {1} {2} MV^{2} \leq  mv^{2} \approx \Delta E.
\end {equation}

\par
In this case, the complete kinetic energy of the projectile is insufficient
for exciting the atom.
We then have only elastic scattering with the atom staying always in the ground
state.$^{16}$
We also take the mass of the nucleus to be so large that its recoil
velocity and recoil energy are completely negligible.
\par
For the rest of this section, we will assume that the projectile-electron
interaction $V_{int}(\vec{r} - \vec{R} (t))$ is weak enough so that the
scattering cross section can be computed in the Born approximation:
\begin {equation}
\frac {d \sigma} {d \Omega}
\approx \mid f_{B} (\theta) \mid ^{2}
\approx  \mid F(\Delta \vec{P}) \mid ^{2} (G(\Delta P))^{2}
\end {equation}
with $F(\Delta \vec{P})$ the form factor defined in Eq. (18) above and $G(\Delta P)$  the interaction potential in Fourier space:
\begin {equation}
G(\Delta \vec{P}) \equiv
\int e^{i \Delta \vec{P} \cdot \vec{\rho}}
V_{int}(\vec{\rho}) d \vec{\rho}.
\end {equation}
(Thus for Coulomb interaction
$G(\Delta \vec {P})
\approx \alpha _{em}/ \mid \Delta P \mid ^{2})$.
By measuring
$d \sigma / d \Omega$,
we can then infer
$F(\Delta \vec {P})$
and
$\psi _{0} ^{2} (\vec{r})$.
Since the atom is never excited, we need just one atom from which we
repeatedly scatter
(a beam of ) many projectiles.
It is amusing to ask how much time is required in such an experiment in
order to map out the ground state wave function to within an accuracy of
10$^{-3}$.
For an incident flux $\Phi$ the rate of collision will be
$dN/dt = \Phi \sigma_{el}$ with $\sigma_{el} \approx \epsilon a_{0} ^{2}$
the elastic cross section.  (The parameter $\epsilon \ll 1$ expresses the
weakness of the interaction.)  In order to obtain an $O(1/N)$ precision in
the determination of
$F(\Delta \vec{P})$
(or $\psi_{0}(\vec {r})$),
we need $N$ scattering.  The ``experiment'' should therefore be carried
over a time $\tau$ such that:
\begin {equation}
\Phi \sigma \tau = N
\end {equation}
Two considerations limit the flux $\Phi = n V$ with n the number density of
projectiles in the beam.  First, from Eq. (38) $V \leq \sqrt {m/M} v = v /
\sqrt {f}$.  Also we take the number density of projectiles $n \leq
a_{0}^{-3}$, to avoid simultaneous scattering of several projectiles.
Thus, we find
\begin {equation}
\tau \geq \frac{a_{0}} {V \eta} \frac {N} {\epsilon}
\geq \frac {Na_{0} \sqrt {f}} {\eta \epsilon v}
= \frac {N \sqrt {f} t_{atom}} {\eta \epsilon}
\end {equation}
with $\eta$ a measure of how far we are from the ``maximal density'' $ n
\approx a_{0} ^{-3}$.  Even for a large $f$, a small $\epsilon$
(guaranteeing the validity of the Born approximation) and a tiny $\eta$:
\begin {equation}
f = M/m = 10^{4}
\end {equation}
\begin {equation}
\epsilon \approx 10^{-3}
\end {equation}
\begin {equation}
\eta \approx 10^{-6}
\end {equation}
We need ``only'' $\tau \simeq 0.01 sec$ in order to obtain
\begin {equation}
N \approx 10^{3}.
\end {equation}
This is due to the very short atomic period, $t_{atom}$, which for
hydrogen, is $10^{-16}$ sec.

{\bf V.     Protective Scattering: The Semi-classical Approximation.}

\par
In the previous section we have shown that scattering of slow massive
projectiles, when treated in the Born approximation, can allow finding the
full form factor $F(\Delta \vec {P})$ and hence the complete wave function
$\psi_{0} (r)$ of the atom, even if only a single atom is available.  While
this demonstrates the measurability of the wave function of a single
system, the scattering considered differ form the protective measurements
of Aharonov and Vaidman and Aharonov, Anandan, and Vaidman$^{3}$ in one
crucial respect.  The outcome of any single scattering was probabilistic
with the probability for scattering with momentum transfer $\Delta \vec{P}$
being proportional to $\mid F (\Delta \vec{P}) \mid^{2}$ -whereas in the AV
and AAV protective measurements, the pointer deflection [i. e. scattering
angle] was deterministic.
\par As will be shown next, the latter, deterministic case, can also be
modeled by semi-classical scattering. To this end we will maintain the
conditions  for protective scattering.  (i)  The kinetic energy of the
projectile satisfies $\frac {1} {2} MV^{2} \leq \Delta E$ so that inelastic
scattering cannot happen, though $MV \rangle \frac {\hbar} {a} \approx mv$.
(ii)  The projectile-electron interaction is weak enough so that the
original wave function is not distorted.  We will however relax the
assumption that scattering can be treated in the Born approximation.
For convenience we will assume a coulombic interaction between the electron
of change q and projectile of change $q '$:
\begin {equation}
V_{I} (\vec {r} - \vec {R}) = \frac {q q'} {\mid \vec{r} - \vec{R} \mid} .
\end {equation}
[Taking q and $q '$ to have the same sign disallows rearrangement
reactions where the electron is picked up by the projectile.]  We really
need not specify the binding potential $V_{B} (r)$.  However, to make the
discussion more concrete, we also take a Coulombic binding potential:
\begin {equation}
V_{B} (\vec {r}) = - \frac {Zq^{2}} { \mid \vec {r} \mid }.
\end {equation}
Our requirement $V_{B} \gg V_{I}$ is implemented by having
\begin {equation}
Zq \gg q '
\end {equation}
\par
To simplify the treatment, we assume that there is no direct
projectile-nucleus interaction.  We assume as in (ii) above that $V_{I}
(\vec {r} - \vec{R})$ constitutes a small perturbation for the \underline
{electron}.  We will \underline {not} assume that it is weak insofar as the
\underline {projectile} motion is concerned. 
The weaker projectile ($X$  particle) -- electron interaction as
compared with electron-proton interaction is conveyed in Fig. 1a by
the larger number of virtual quanta (photons exchanged) between $e$
and $P$.
\vskip 0.2cm
\epsfysize=8 cm
 \centerline{\epsfbox{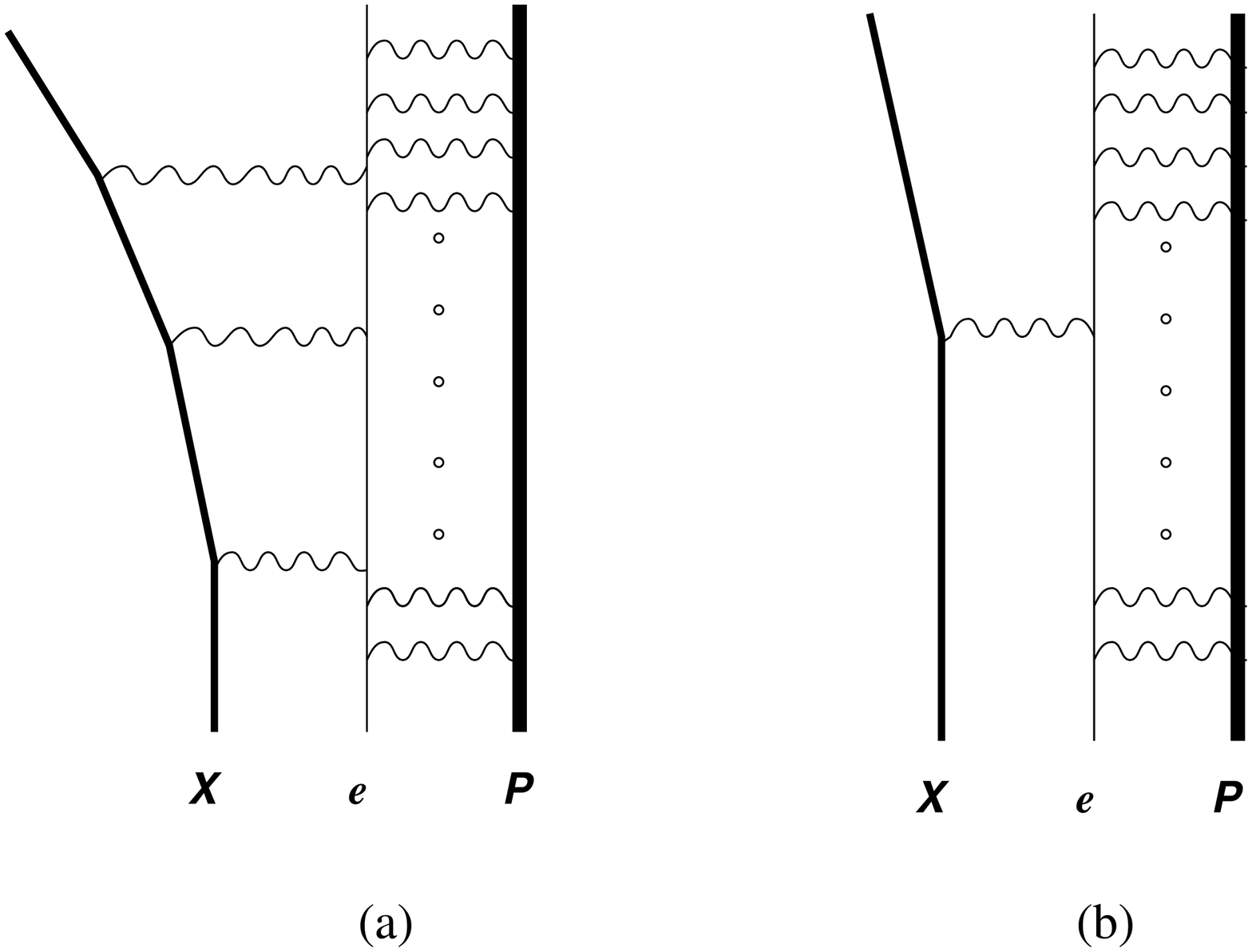}} 
{\small {\bf Figure 1}: {\footnotesize 
 Scattering of a heavy projectile $X$ on the light electron $e$
bound strongly to a super heavy proton $P$. The thickness of
the lines conveys the information on the masses of the particles.~
(a) The scattering  in a
semi-classical, many photon exchange, approximation.~(b) The scattering in the Born, one photon exchange, approximation.
}}
\par
The phase picked up by the projectile as it moves along a straight line
parallel to the z axes at an impact parameter with fixed velocity
$V:\vec{R}(t) = \vec{b} + Vt  \hat{e}_{z}$ is approximately given by:
\begin {equation}
\phi \approx \frac {1} {\hbar} \int_{path} dtV_{I} (\vec {r} - \vec{R} (t))
= \frac {q q'} {\hbar V} ln (\frac {\Lambda}{b})
\end {equation}
with $\Lambda$ a cut-off length.
By having
\begin {equation}
\frac {qq'} {\hbar V} \equiv \eta \gg 1
\end {equation}
we make the projectile electron interaction strong.
The large Coulomb phase and the large angular momentum
\begin {equation}
L = \frac {MVb} {\hbar} \approx \frac {MVa} {\hbar} \approx \frac {MV} {mv}
\gg 1
\end {equation}
suggest that we can treat the projectiles' motion semi-classically.
The Feynman diagrammatic expression of this is the fact that the momentum
transfer $\Delta P$ is built up by exchanging a large number $n_{\gamma}
\approx \eta \gg 1$ of photons, see Fig. 1a, rather than via a single
photon exchange as in the case  of the Born approximation, see Fig 1b .
A semi-classical computation of the deflection, assuming still, for
convenience, that $\theta$-the scattering angle is small, can be done as
follows.  The electrostatic force exerted by the (spherically symmetric!)
charge distribution of the electron on the projectile is
\begin {equation}
\vec{F} = \frac{Q(R) q ' \hat{e}_{b}} {R^{2}}
\end {equation}
where Q(R) is the total electric charge within a sphere of radius $R = \mid
\vec{R} (t) \mid$
\begin {equation}
Q(R) = 4 \pi q \int_{0}^{R} r^{2} dr (\psi_{0}(r))^{2}.
\end {equation}
[We note that the use of the average value (for the charge inside R) in the
state $\psi_{0}$ parallels the general discussion in Section II above where
the average
\newline
$\langle \psi_{0} \mid A \mid \psi_{0}\rangle$ determined the deflection of the pointer.]
The momentum transfer is then:
\begin {equation}
\Delta \vec{P} = \int_{-\infty}^{\infty} \vec{F} dt = q'
\int_{-\infty}^{\infty} \frac {dz} {V} \frac {Q(R=\sqrt {z^{2}} + b^{2})}
{(z^{2} + b^{2})^{3/2}} b \hat{e}_{b}
\end {equation}
where we used $dt = dz/V$ and the fact that the net momentum transfer in
this approximation [with the projectile moving with uniform velocity
parallel to the $z$ axes] is purely transverse i. e. along $\hat{e}_{b}$.
The order of magnitude of $\Delta P$ is readily estimated:
\begin {equation}
\Delta P \approx F \delta t_{col} \approx \frac {1} {\hbar} \frac
{qq'}{a_{0}^{2}} \frac {a_{0} \hbar} {V} = \eta \frac {\hbar}{a_{0}}.
\end {equation}
Such a $\Delta P$ is indeed expected if, as conjectured before, we have
$n_{\gamma} \approx \eta$ photons exchanged, each contribution a typical
momentum transfer of $= \hbar/ a_{0} \approx mv$.
This multiple exchange make the total momentum transfer larger by a factor
$\eta$ and easier to measure.  Also the large number of quanta involved
tends to make the deflection almost deterministic so that the above
classical treatment of the scattering is justified.
[Parenthetically we note that a small deflection:
\begin {equation}
\theta = \frac {\Delta P} {P} = \frac {\eta m v} {MV} < 1
\end {equation}
requires a stronger version of Eq. (36).
\begin {equation}
MV > \eta m v; \eta \equiv \frac {qq'}{\hbar V} \gg 1
\end {equation}
While the last equation implies that
\begin {equation}
MV^{2} > \frac {qq' v m} {\hbar} \approx \frac {qq'} {a_{0}}
\end {equation}
the latter does not conflict with the basic requirement
\begin {equation}
kE = \frac {1} {2} MV^{2} < \Delta E = \frac {Zq^{2}} {a_{0}}
\end {equation}
proving that the ratio $Zq/q'$ is large enough.  This however is
precisely our Eq. (49) above.]
\par
To measure the wave function, it suffices to find Q(r) of Eq. (54) since
$\psi_{0}(r)$ is given by:
\begin {equation}
\psi_{0} (r) \approx \frac {1} {r} \sqrt{\frac{dQ(r)}{dr}}.
\end {equation}
Scattering projectiles with various initial impact parameters b and
measuring in each case the momentum transfer $\Delta P (b)$ given via Eq.
(55), we can solve for $Q(r)$. [In practice also such classical scattering,
where $\Delta P$ is uniquely determined for a given b by a calculable
function $\Delta P (b)$, are carried out by using an incident uniform beam
of projectiles and measuring $d\sigma/ d \Omega$.$^{17}$]

\vskip 0.2cm

{\bf VI.     Summary and Discussion.}

\par
The arguments of this paper, complementing and slightly extending the
original work of AAV and AV$^{3}$ definitely show that there is no quantum
mechanics subtlety hinders the measurement of the wave function of a single
atom.  In all cases we require a projectile which is much heavier than the
particle say electron whose wave function is being measured.
The set-ups considered were still somewhat idealized.  Thus in Section V we
assumed that there is a nucleus-electron and electron-projectile
interaction, but no direct interaction between projectile and nucleus.
These interactions  could be Coulomb-like only if the electron carried two
types of charges: $q \epsilon U (1)$, and $q' \epsilon U' (1)$
sharing with the nucleus one type of charge and with the projectile the
other type.  While the absence of projectile nucleus interaction is not an
absolute must, it excludes the possibility of the projectile being captured
around the nucleus.  Ideally, the projectile should also:
\newline
(a) be point like;
\newline
(b) have long lifetimes;
\newline
(c) have minimal interaction with the nucleus
\newline
Clearly (a) - (c) are a ``convenience'' issue.  Thus in principle we could
use whole atoms or ions as massive slow projectiles.  The separation of the
target wave function from the effects due to the structure of the
projectile, would then be more difficult.
The muon is point like and $m_{\mu} = 200 m_{e}$.  However, to satisfy the
basic condition (37):
\begin {equation}
E_{\mu} = \frac {1} {2} m_{\mu} \beta_{\mu}^{2} \leq \Delta E_{Hydrogen}
\approx 10 eV
\end {equation}
we need very slow muons.  The muons' lifetime is only $~2.10^{-6}$ seconds
and these slow muons propagate only a few centimeters before decaying.  The
task of generating intense, monochromatic, collimated muon beams,
scattering them and later, detecting the scattered muons seems very
difficult.
\par We could also utilize neutrons which are heavy and fairly long lived.
However, the low energy neutrons $(E_{n} \leq 10 eV)$ often have strong
nuclear interactions which could mask the effects of the electron-neutron
scattering.
\par
Amusingly, the scattering of hypothetical WIMPs (``Weakly Interacting
Massive Particles'') of mass, say $200 GeV$ on heavy nuclei $(A \geq 200)$,
provides a simple example of a form factor appearing when the interaction
is purely elastic.  For $M \approx 200 GeV$, the kinetic WIMP energy:
$E_{WIMP} = \frac {1} {2} MV^{2} \approx 100 KeV$ (for $V \approx
10^{-3}c$, typical virial velocity) is below nuclear excitation. Thus our
basic no excitation requirement (Eq. (38a)) is readily satisfied.  At the
same time the maximal momentum transfer $\Delta P \approx 200 MeV$ is
appreciable and $\Delta P R(A; Z) \gg 1$ so that $F(\Delta P)$ could be
significantly smaller than unity.
\par
In conclusion, it is amusing to note that quantum systems are actually more
rigid than classical systems.  The latter have continuous spectra and can
suffer arbitrarily small excitation.  Any scattering from a classical
target would leave the latter slightly deformed.  This however is not the
case for quantum mechanical system

\vskip 0.2cm

{\bf Acknowledgments.}

I have benefited from hearing first-hand about the measurement of the wave
function from the authors of the original papers (Y. Aharonov, J. Anandan,
and L. Vaidman).  The trigger for the present work, beside the desire to
have more concrete realization of their ideas, was a comment by F. T.
Avignone concerning the form factors for scattering of WIMPS for which I am
particularly grateful. The author would like to acknowledge the USA  -- Israel Science Foundation.

\break

\end {document}